**PAPER**

Check for updates

# Convergence acceleration in machine learning potentials for atomistic simulations†

Dylan Bayerl,[a] Christopher M. Andolina, [a] Shyam Dwaraknath [b] and Wissam A. Saidi *[a]

Machine learning potentials (MLPs) for atomistic simulations have an enormous prospective impact on materials modeling, offering orders of magnitude speedup over density functional theory (DFT) calculations without appreciably sacrificing accuracy in the prediction of material properties. However, the generation of large datasets needed for training MLPs is daunting. Herein, we show that MLP-based material property predictions converge faster with respect to precision for Brillouin zone integrations than DFT-based property predictions. We demonstrate that this phenomenon is robust across material properties for different metallic systems. Further, we provide statistical error metrics to accurately determine *a priori* the precision level required of DFT training datasets for MLPs to ensure accelerated convergence of material property predictions, thus significantly reducing the computational expense of MLP development.

## Introduction

First-principles-based simulations at the atomic level have made significant strides in the past decade designing novel materials for batteries to microelectronics.[1–4] Unfortunately, the state-of-the-art first-principles methods, such as density functional theory (DFT), are limited to a few nm³ in volume, on the order of hundreds of atoms, and time scales on the order of nanoseconds. On the other hand, industry is very adept at modeling the continuum mechanics level: from μm³ and microseconds on up. Thus, there is a *valley of death* in scaling, a challenge that should, in theory, be addressable by current atomistic modeling methods such as molecular dynamics. In fact, the 2013 Nobel prize in chemistry was awarded for the development of combined quantum mechanics/molecular dynamics modeling to bridge the *valley of death*.

Central to the success of molecular dynamics is the existence of force-fields or atomic potentials with sufficient fidelity to describe atomistic interactions. For instance, the embedded atom method (EAM) is widely used to describe metal–metal interactions.[5] Bond order potentials such as Brenner,[6] Tersoff,[7] and Stillinger–Weber[8] are used to describe covalently bonded systems. The Buckingham potential was introduced to describe the short-range interactions in ionically-bonded systems, while Coulomb interactions are used for long-range electrostatic interactions.[9] Reactive force-fields (ReaxFF)[10,11] and charge-optimized many-body (COMB)[12] can describe complex atomistic interactions such as charge transfer and bond breaking. While indispensable for large-scale atomistic modeling, these atomistic force-fields have significant limitations including non-transparency, time-intensive development, limits on accuracy and transferability to different properties, and being unavailable for many material systems. As an alternative to traditional force-field-based atomistic potentials, machine learning-based potentials and particularly deep neural networks have demonstrated the flexibility necessary to model the complex potential energy surfaces of atomistic interactions.[13–19]

Machine learning potentials (MLPs) allow simulating atomistic interaction energies and forces without any explicit functional form, distinct from the specific functional interactions of conventional potentials.[20–24] However, this freedom comes at the cost of requiring a large dataset typically generated by DFT for training. For instance, the Behler group developed a Cu MLP using 35k configurations[25] and a model for CuZnO using 100k configurations.[26] The recent GAP Fe potential developed by Csányi and Marzari utilized approximately 150k local atomic environments, although this MLP can only describe the solid phase.[27] The Saidi group developed versatile binary alloy MLPs for Cu–Zr, Al–Mg, and Au–Ag that can describe many properties for these systems through MLP-based molecular dynamics simulations requiring, respectively, 302k, 250k, and 85k configurations.[28–30] Reducing the size of the training set for developing MLPs is an active field of research.[31,32] However, the large amount of training data needed to develop a single MLP poses a significant limitation on utilizing this new frontier of research in materials modeling, which urgently calls for the development of efficient computational frameworks for generating training datasets.

[a] *Department of Mechanical Engineering and Materials Science, University of Pittsburgh, Pittsburgh, Pennsylvania 15216, USA. E-mail: alsaidi@pitt.edu*

[b] *Energy Technologies Area, Lawrence Berkeley National Laboratory, 1 Cyclotron Road, Berkeley, California 94720, USA*

† Electronic supplementary information (ESI) available: Table of conversions between measures of *k*-space sampling density for investigated metal lattices. Parity plots of force and virial predictions from MLPs and DFT. Plots of all studied material properties for all studied material systems as functions of DFT precision (*k*-space sampling density) calculated by DFT and MLPs. Plots of all 4 statistical error metrics for the DFT test dataset as functions of precision. Input files for AIMD simulations in VASP and MLP training with DeepMD-kit. For input files and downloadable datasets of calculated property data, see http://d-scholarship.pitt.edu/id/eprint/41716. See DOI: 10.1039/d1dd00005e

In standard DFT calculations, two main parameters significantly impact the computational cost, precision, and convergence of accurate property predictions. One is the size of the basis functions needed to expand the Kohn–Sham orbitals, which is closely related to how electron–nucleus interactions are described. For example, in plane-wave calculations, the cutoff energy controls the number of plane-waves in the basis set with the energy of the system converging variationally with the size of the basis. The second is the $k$-space sampling density of the first Brillouin zone (BZ) of the system. The appropriate number of $k$-points for a DFT calculation depends on the material system, the properties of interest, and the level of convergence desired. Convergence procedures for various properties are well-established in the DFT community, and several studies investigated non-traditional $k$-space sampling schemes for standard DFT calculations.[33–35] However, there has been no prior systematic study of the impact of $k$-space sampling density on MLPs trained using DFT datasets. This work aims to bridge this gap and establish guidelines for converging the $k$-space sampling density of training datasets for MLPs.

Herein, we use $k$-space sampling density to control the precision of BZ integrations in DFT calculations and study the impact of variable precision on MLPs. Specifically, we investigate the convergence of bulk Al, Cu, and Mg material properties in BCC, FCC, and HCP lattices calculated directly by DFT and indirectly by MLPs trained on DFT datasets. The overall workflow of the study is summarized in Fig. 1. Our investigation revealed a phenomenon of accelerated convergence of material property predictions by MLPs with respect to the precision of DFT training data. Specifically, even when trained on low-precision DFT data with low sampling density, MLPs can predict properties, energies, and forces consistent with high-precision, high-sampling-density DFT calculations. Further, we provide statistical error metrics to accurately determine *a priori* the precision level required of DFT training datasets to ensure accelerated convergence of MLP property predictions. Our findings have significant implications for DFT-based MLPs, suggesting that the computational cost of training data production can be significantly reduced without sacrificing the accuracy of property predictions.

## Methods

### DFT calculation parameters

We performed DFT calculations in VASP using the Perdew–Burke–Ernzerhof (PBE) exchange–correlation functional with projector-augmented wave pseudopotentials.[36–38] Following the methodological approach of the Materials Project,[39,40] we selected non-GW pseudopotentials with the most available semicore states from the VASP pseudopotential database. We selected a plane-wave cutoff of $1.3\times$ the largest ENMAX among these pseudopotentials. This plane-wave cutoff of 524 eV was used consistently across all materials. We used a tight break condition of $1 \times 10^{-7}$ eV free energy change between steps in the electronic relaxation loop. Moreover, we applied Methfessel–Paxton[41] smearing of 2nd-order with 0.15 eV broadening to electronic occupations. Tests varying the smearing parameter from 0.05 to 0.25 eV indicate little impact (less than 4% change) on mechanical and defect properties that are ultimately the focus of this work. These data, as well as DFT input parameter files are available online in the ESI.†

### DFT precision

In our DFT calculations, we controlled the precision of the calculation by varying the linear density of $k$-space sampling ($\Delta k$) in the first BZ. $\Delta k$ defines the minimum spacing in units of $\text{Å}^{-1}$ between adjacent $k$-points in gamma-centered Monkhorst–Pack grids.[42] Note that $\Delta k$ is one of several methods to describe $k$-space sampling density. Common alternatives include grid notation and specifying the number of $k$-points per reciprocal atom (pra).[43] For example, in the Cu FCC conventional cell, $\Delta k = 0.18 \text{ Å}^{-1}$ sampling yields a $10 \times 10 \times 10$ grid, equivalent to 4000 $k$-points pra. Conversions between $\Delta k$, grid notation, and pra for the studied primitive-cells are tabulated in Table S1.† All AIMD and DFT calculations were replicated at 7 distinct values of $\Delta k$ ranging from $0.12 \text{ Å}^{-1}$ to $0.96 \text{ Å}^{-1}$ to elucidate the impact of DFT precision on property prediction. Note that the upper limit of $\Delta k = 0.96 \text{ Å}^{-1}$ corresponds to gamma-point-only sampling in the majority of $2 \times 2 \times 2$ supercells of our materials. Moreover, we performed additional DFT property calculations at $\Delta k = 0.09 \text{ Å}^{-1}$ to provide an ultimate reference point of extremely high precision for comparison against calculations prepared with larger $\Delta k$ (*i.e.* lower precision).

### Training dataset parameters

To generate data to train MLPs, we performed *ab initio* molecular dynamics (AIMD)-based local-configuration sampling using VASP to map out the potential well around each configuration generated by DP-GEN.[44] Each structural configuration was evolved within the NVE ensemble for 100 steps using a 2 fs timestep starting from initial ion velocities conforming to the Maxwell–Boltzmann distribution at 500 K that is much smaller than the melting temperature for all metals. Initial structural configurations were generated by first relaxing primitive cells of each of the Al, Cu, and Mg pure metals in BCC, FCC, and HCP lattices. Each relaxed primitive cell was then converted to a $2 \times 2 \times 2$ supercell, to which we applied linear scaling factors of 0.96, 0.98, 1.00, 1.02, and 1.04 to all lattice vectors, yielding 5 distinct volumetric deformations. We further applied 2 different random perturbations of up to 3% to each lattice vector and up to 0.15 Å to each ion from its equilibrium position to each volumetric deformation, yielding 10 initial configurations of $2 \times 2 \times 2$ supercells for each lattice. For each volumetric deformation, we also constructed point-defect configurations containing either a vacancy, a tetrahedral ($T_d$) self-interstitial, or an octahedral ($O_h$) self-interstitial, yielding 15 additional $2 \times 2 \times 2$ supercells per lattice. Using the same scale-and-perturb approach, we also generated 13 configurations based on the primitive cells for each lattice, spanning a range of lattice vector rescaling factors from 0.88 to 1.12. Therefore, the complete set of AIMD initial configurations contained $10 + 15 + 13 = 38$ structures with perturbed and scaled ions and lattice vectors for each metal and lattice combination. We replicated this structure generation procedure and performed AIMD calculations independently for each of the 7 values of $\Delta k$ mentioned previously, resulting in 38 structures $\times$ 3 lattices $\times$ 100 timesteps = 11 400 configurations sampled per metal, per $\Delta k$ value. Thus, we generated 7 parametrically identical training datasets, consistent with 7 independent studies utilizing a target $\Delta k$ and developing an MLP.

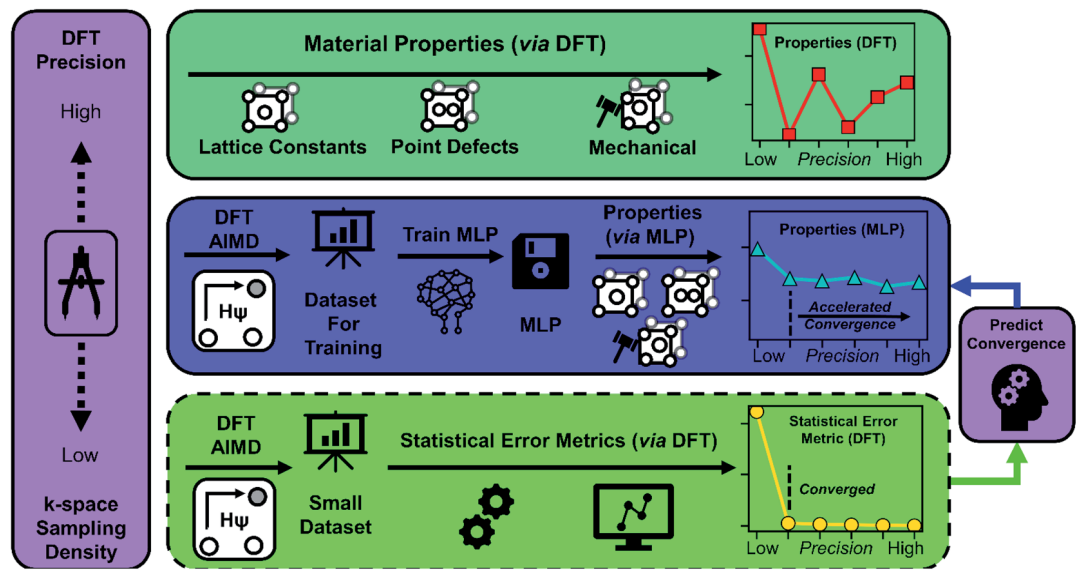

**Fig. 1** The workflow of our investigation into the impact of $k$-space sampling density and corresponding DFT precision on the convergence of material property predictions. At variable levels of precision ($k$-space sampling density), we calculated properties directly by DFT (top row) and through neural network-based MLPs (center row). Properties predicted through MLPs exhibit accelerated convergence over DFT with respect to the precision ($k$-space sampling density). Additionally, analysis of small datasets of DFT calculations with statistical error metrics (bottom row) are capable of predicting the precision regime where MLPs exhibit accelerated convergence.

### MLP training protocol

We trained MLPs on our AIMD-generated datasets using DeepMD-kit[45] within the DeepPot-SE[46] approach. DeepMD-kit utilizes neural networks to interpolate the relationship between atomic coordinates (model input samples) and the energies, forces, and virials (model output labels) in DFT training data. However, we expect our findings to apply to other ML approaches for generating MLPs. Following the generation of separate training datasets for each $\Delta k$ value, we independently trained MLPs on the 11 400-configuration dataset for each metal at each $\Delta k$, yielding 3 metals $\times$ 7 $\Delta k$ values = 21 MLPs. We used a consistent training protocol with identical hyperparameters for each MLP, including randomly initialized weights and 3 layers of 240 fully-connected nodes with hyperbolic tangent activation functions in the neural networks. The complete set of hyperparameters and network architecture details we used for training are provided in a DeepMD-kit input file in the electronic dataset accompanying this work,[47] which we applied sequentially 4 times by the warm-restart method[48] to train each MLP.

### Material property calculations

To evaluate the performance of our MLPs, we tested their predictions of material properties against DFT calculations. In particular, we compared MLP property predictions to DFT calculations at the same $\Delta k$ value as the underlying training dataset for each MLP. With all datasets and models generated and trained in a fully-consistent and parametrically identical fashion, this analysis can reveal how the DFT precision (as controlled by $\Delta k$) propagates from the training dataset to MLP predictions.

We calculated a variety of bulk mechanical and defect properties in each metal and lattice. Specifically, we calculated the Eulerian–Birch[49] equation of state (EOS)-derived quantities such as equilibrium volume-per-atom and cohesive energy; linear-elastic mechanical properties including stiffness matrix elements, the Poisson ratio, as well as bulk, shear, and Young's moduli; and point defect formation energies for vacancies, $O_h$, and $T_d$ self-interstitials. The equations and procedures for these property calculations have been fully described elsewhere.[28] Moreover, since linear-elastic mechanical property calculations can be sensitive to the magnitude of strain applied to the simulation cell, we calculated average values of these properties for normal deformations ranging from 1% to 3% and shear deformations ranging from 3% to 7%. We used VASP for DFT

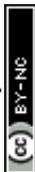

calculations and the DeepMD-kit-compatible version of LAMMPS[50] for MLP calculations.

## Results and discussion

To verify that our MLPs are effectively learning (*i.e.*, accurately interpolating) the DFT training data, we tested the prediction parity of MLP predictions *versus* DFT training data across all 21 MLPs (Fig. 2). The energy-per-atom predictions of MLPs generally exhibit good agreement with DFT (Fig. 2a), suggesting that our model hyperparameters and training protocol result in effective learning. Notably, as DFT training data precision is decreased by increasing the *k*-space sampling density parameter $\Delta k$, the MLP energy-per-atom predictions exhibit more significant deviations from parity. Importantly, we do not see systematic deviations from the DFT values for the 7 different $\Delta k$ values. Similar behavior is observed in the force and virial component predictions (Fig. S1†), though force and virial errors increase slightly faster than energy error with increasing $\Delta k$, as is expected. This observation is quantified in Fig. 2b, showing the root mean squared error (RMSE) of MLP predictions of the energy-per-atom, force-components, and virial-components-per-atom against DFT as a function of $\Delta k$. The increasing RMSE with $\Delta k$ suggests that the models are not overtraining on the increasingly noisy, lower-precision DFT data. This observation is further supported by comparing the training data RMSE (Fig. 2b) to the RMSE of a test dataset of DFT calculations unseen by the MLPs during training (Fig. 2c). Significantly larger RMSE in the test *versus* training dataset would indicate overtraining of MLPs, such that they do not generalize to unseen data. However, we observe similar RMSE as a function of $\Delta k$ in both training and test data, indicating that our models generalize well to novel atomic configurations.

### Impact of precision on the convergence of material properties

The parity plot of Fig. 2 shows clearly that the fidelity of the MLPs deteriorates with the decrease in the precision of the training dataset. However, it is important to assess how this affects the material properties predicted by the MLPs for practical applications. We investigate bulk mechanical, EOS, and point defect properties, which emerge from the collective interactions of multiple atoms. DFT and MLP predictions of the vacancy formation energy are shown as an example in Fig. 3, while the additional properties considered in this work are shown in Fig. S2.† Consider first the DFT-predicted values in each metal for each lattice. It is evident by inspection that DFT predictions exhibit considerable variation as $\Delta k$ increases, such as in FCC-Al. Indeed, in some cases, the DFT-prediction varies by more than 50%, with a slight change in $\Delta k$. While DFT-predictions appear to converge as $\Delta k$ decreases (precision increases), it is not necessarily clear which $\Delta k$ should be chosen to converge all property predictions within a predetermined accuracy without first calculating all properties of interest over a wide range of $\Delta k$.

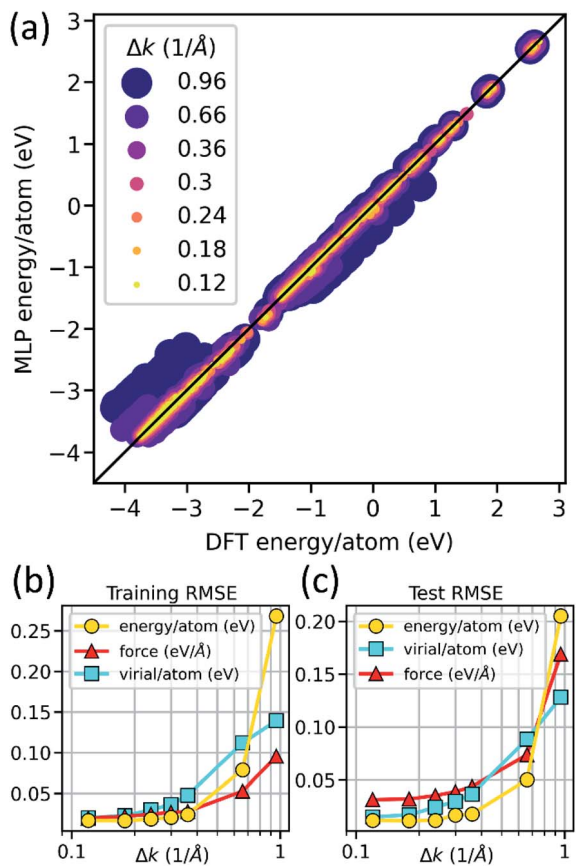

**Fig. 2** Parity plot of energy per atom from MLP predictions and DFT calculations for all models and training data generated in this study (a). Legend indicates *k*-point sampling density ($\Delta k$) of the DFT training data. MLP deviation from parity increases with decreasing precision of the DFT training dataset (increasing $\Delta k$). The RMSE quantifies deviation as a function of $\Delta k$ in training (b) and test (c) datasets for energy per atom, virial components per atom, and force components per atom. Similar RMSE in both the training and test datasets show that the MLPs are not overtrained.

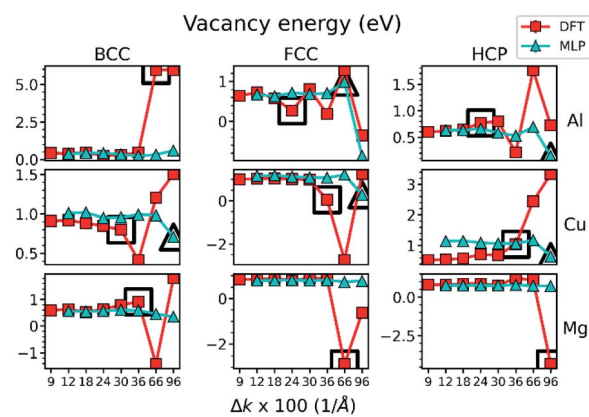

**Fig. 3** The vacancy formation energy of Al, Cu, and Mg metals in BCC, FCC, and HCP lattices predicted by DFT calculations and MLPs as a function of $\Delta k$. Threshold $\Delta k$ ($\Delta k_T$) where property prediction deviations from that at $\Delta k_{min}$ by $\frac{1}{2}$MAD$_{max}$ are highlighted with prominent, black-outlined symbols. The $\frac{1}{2}$MAD$_{max}$ criterion demarcates the $\Delta k$ value below which property prediction is stable, and is defined in detail in the main text. $\Delta k_T$ is generally smaller for DFT than MLPs, demonstrating the higher stability and accelerated convergence of MLP predictions.

Next, consider the MLP-predicted values of the vacancy energy in Fig. 3. The most notable feature is their stability with respect to the $\Delta k$ used to generate the training data. In contrast with DFT predictions, the MLP-predicted vacancy energy varies negligibly below an apparent threshold $\Delta k$ value (*e.g.*, Cu, FCC- and HCP-Al), or in some cases over all $\Delta k$ (*e.g.*, Mg and BCC-Al). Similar behavior is observed in other properties (Fig. S2†). Below a threshold $\Delta k$ value ($\Delta k_{\mathrm{T}}$), MLP predictions are generally very close to the value calculated at the minimum $\Delta k$ ($\Delta k_{\mathrm{min}}$). In this sense, the convergence with respect to $\Delta k$ is accelerated below $\Delta k_{\mathrm{T}}$ in MLPs.

We adopt the mean absolute deviation (MAD) as a comparative measure of property prediction variability to quantify the apparent convergence acceleration and stability of MLPs below $\Delta k_{\mathrm{T}}$. The MAD is expressed as $\mathrm{MAD} = \frac{1}{N} \sum_i^N |x_i - \bar{x}|$ where $x_i$ are the property values for each $\Delta k$ and $\bar{x}$ is the average value of the property over $N$ values in the range of $\Delta k$ included in the summation. For example, in FCC-Al the vacancy energy for $\Delta k \leq 0.96$ Å$^{-1}$ has MAD of 0.174 eV from MLP and 0.519 eV from DFT (Table 1). The consistently larger MAD of DFT *versus* MLP quantitatively confirms the larger variation in DFT than MLP (Fig. 3). Considering just $\Delta k < \Delta k_{\mathrm{T}}$ to evaluate the MAD does not change the conclusion. For instance, if we recalculate the MAD for $\Delta k < 0.66$ Å$^{-1}$ for FCC-Al, we obtain 0.024 eV for MLP and 0.206 eV for DFT. Quantified in this way, the variation of the FCC-Al vacancy energy below the MLP threshold $\Delta k$ of 0.66 Å$^{-1}$ is approximately 9 times lower in MLP than DFT.

Alternatively, consider the maximum deviation of the FCC-Al vacancy energy within $\Delta k < \Delta k_{\mathrm{T}}$ from the ultimate prediction at $\Delta k_{\mathrm{min}}$. The largest deviation of MLP from the ultimate value of 0.671 eV is +0.045 eV (0.716 eV at $\Delta k = 0.24$ Å$^{-1}$), whereas the largest deviation of DFT from its ultimate value of 0.641 eV is −0.376 eV (0.265 eV at $\Delta k = 0.24$ Å$^{-1}$). By this measure, the MLP predictions vary approximately 8 times less than DFT, but the MAD is a more convenient global measure of variability than maximum deviation and will be used from here on. An identical analysis of the MAD over $\Delta k$ for other properties in Fig. S2† shows that our findings for the vacancy energy (Table 1) of convergence acceleration and improved property prediction stability in MLPs *versus* DFT generally holds for the other properties considered in this work.

### Threshold $\Delta k$ criterion demarcating convergence acceleration

As we have shown, the MAD quantitatively captures the convergence acceleration and property prediction stability of MLPs that is intuitive from graphical comparison to DFT (Fig. 3 and S2†). We then utilized the MAD to formulate a scale-invariant criterion for determining $\Delta k_{\mathrm{T}}$. Scale invariance is convenient as it would be independent of the magnitude of the dimensionalized units for all the different material properties under consideration. We propose a $\Delta k_{\mathrm{T}}$ criterion based on,

$$\min_{\Delta k} |x(\Delta k) - x(\Delta k_{\mathrm{min}})| \geq \frac{1}{2}\mathrm{MAD}_{\mathrm{max}} \qquad (1)$$

where $x(\Delta k)$ is the MLP or DFT property prediction at $\Delta k$, and $\mathrm{MAD}_{\mathrm{max}}$ is the larger of the DFT and MLP MAD over all $\Delta k$ for a given property, metal, and lattice (*e.g.*, $\mathrm{MAD}_{\mathrm{max}} = 0.519$ eV for the vacancy energy of FCC-Al, from Table 1). Using the larger MAD value from DFT *versus* MLP prevents erroneous selection of $\Delta k_{\mathrm{T}}$ within the convergence-accelerated regime when the MAD is very small (*i.e.*, the property prediction is stable over all $\Delta k$), such as in the MLP vacancy energies of Mg (see Fig. 3 and Table 1). This criterion for $\Delta k_{\mathrm{T}}$ is supported by the trend of large black-outlined symbols in Fig. 3 and S2.†

### Impact of precision on statistical properties of DFT datasets

To elucidate the mechanism of accelerated convergence of MLP material properties, we analyzed properties of DFT training data as a function of $\Delta k$. Since the training process is statistical, we focused on analyzing the overall statistical properties of DFT data. Moreover, since MLPs ultimately predict material properties by inferencing energies, forces, and virials of atomic configurations, we focused on analysis related to these quantities.

For this purpose, we prepared a distinct test dataset (entirely separate from MLP training data) of 3600 structural configurations on which we performed single-point DFT calculations of energies, forces, and virials as a function of $\Delta k$ in the range from 0.09 Å$^{-1}$ to 0.96 Å$^{-1}$ specified previously. This test dataset simulates a small sub-sample of an MLP training dataset and contains 100 variously scaled and perturbed configurations for each of the 36 combinations of metal, lattice, and structure type (pristine, vacancy, $T_d$, and $O_h$ interstitial). Crucially, single-point calculations were performed on precisely the same 3600 configurations for each $\Delta k$ value, such that any variation in energy, forces, or virial of any individual configuration is due solely to changes in $\Delta k$. We expected the numerical noise introduced by increasing $\Delta k$ to have a measurable impact on the statistical properties of the training data, ultimately giving rise to instability in property prediction at $\Delta k_{\mathrm{T}}$.

We used 4 statistical error metrics (SEMs) to analyze statistical properties of the test datasets: the averaged energy-per-atom deviation (AEAD), the energy-per-atom RMSE (E-RMSE), the RMSE of each force component on each atom (F-RMSE),

**Table 1** The vacancy energies in eV at $\Delta k_{\mathrm{min}}$ from DFT (upper line) and MLP (emboldened lower line) and the mean absolute deviation (MAD) of the vacancy energy over all $\Delta k$

| DFT ± MAD | | | |
|---|---|---|---|
| **MLP ± MAD** | BCC | FCC | HCP |
| Al | $0.43 \pm 2.79$ | $0.64 \pm 0.52$ | $0.60 \pm 0.42$ |
| | $\mathbf{0.37 \pm 0.07}$ | $\mathbf{0.67 \pm 0.17}$ | $\mathbf{0.63 \pm 0.07}$ |
| Cu | $0.91 \pm 0.28$ | $1.01 \pm 1.40$ | $0.53 \pm 1.10$ |
| | $\mathbf{1.01 \pm 0.04}$ | $\mathbf{1.17 \pm 0.10}$ | $\mathbf{1.15 \pm 0.06}$ |
| Mg | $0.59 \pm 0.93$ | $0.84 \pm 1.30$ | $0.79 \pm 0.95$ |
| | $\mathbf{0.56 \pm 0.04}$ | $\mathbf{0.78 \pm 0.01}$ | $\mathbf{0.75 \pm 0.01}$ |

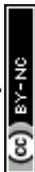

and the RMSE of each virial component per number of atoms (V-RMSE). In each case, the DFT calculations at $\Delta k_{min} = 0.09$ Å$^{-1}$ were used as reference values, such that the SEMs are precisely zero at $\Delta k = \Delta k_{min}$. Furthermore, we calculated each SEM separately for groups of configurations with the same metal, lattice, and structure type (pristine, vacancy, T$_d$, and O$_h$ interstitial). Finally, we imposed the same threshold criterion on material properties to define $\Delta k_T$ for each SEM. For example, the E-RMSE SEM as a function of $\Delta k$ with its calculated $\Delta k_T$ values is shown in Fig. 4, while the remaining SEMs are shown in Fig. S3.†

From inspection of Fig. 4 and S3,† it is evident that these SEMs exhibit qualitatively similar behavior to MLP material property predictions. In particular, SEM magnitudes are uniformly small below $\Delta k_T$, just as MLP variation is low in the convergence-accelerated regime below their respective $\Delta k_T$ values. However, the specific behavior differs subtly between the different SEMs. For example, the E-RMSE (Fig. 4 and S3b†) is generally very stable (near zero) below its $\Delta k_T$ values that are always larger than 0.36 Å$^{-1}$, while the V-RMSE (Fig. S3d†) exhibits a gradual increase with increasing $\Delta k$ and has $\Delta k_T$ values as small as 0.24 Å$^{-1}$. However, the overall behavior suggests that $\Delta k_T$ of these SEMs ($\Delta k_T^{SEM}$) may correlate with $\Delta k_T$ of MLP material properties ($\Delta k_T^{MLP}$) in Fig. 3 and S2.† If so, this would enable *a priori* determination of appropriate $\Delta k$ for DFT training data that is needed to achieve the accelerated convergence of property prediction in MLPs. This method could be utilized when developing MLPs for new material systems to save considerable time and computational resources.

## Predicting the precision threshold for accelerated convergence

Next, we evaluate the ability of the SEMs to predict $\Delta k_T$ of MLP material properties. For each metal and lattice type, we compared $\Delta k_T^{MLP}$ to $\Delta k_T^{SEM}$. Three outcomes are possible. If $\Delta k_T^{MLP} = \Delta k_T^{SEM}$, the SEM deviates by the threshold criterion at the same $\Delta k$ as the material property. If $\Delta k_T^{MLP} > \Delta k_T^{SEM}$, the SEM deviates at smaller $\Delta k$ than the material property, and *vice versa* for $\Delta k_T^{MLP} < \Delta k_T^{SEM}$, indicating mismatch in prediction.

The most important indicator of how well each SEM predicts $\Delta k_T^{MLP}$ is the % mismatch. A larger % mismatch means that $\Delta k_T^{SEM}$ is higher than $\Delta k_T^{MLP}$ for a larger fraction of lattices, structures, and properties. In these cases, the SEM incorrectly predicts convergence acceleration at larger $\Delta k$ than in MLP. Selecting $\Delta k$ for training data that is too large could result in an MLP that cannot accurately predict properties using accelerated convergence. Another important indicator of the predictive capability of each SEM is the % conservative. This is the proportion for which $\Delta k_T^{SEM}$ is lower than $\Delta k_T^{MLP}$, where the SEM conservatively predicts convergence acceleration at $\Delta k$ below where it occurs in MLP. A large % conservative indicates that the SEM is more likely to predict $\Delta k$ for training data yielding convergence-accelerated MLPs, but perhaps with $\Delta k$ smaller (more computationally expensive) than is optimal. The proportion of optimal predictions is captured by the % exact, which indicates where $\Delta k_T^{SEM}$ exactly matches $\Delta k_T^{MLP}$. The exact prediction is desirable since it identifies $\Delta k$ that is no more computationally expensive than necessary to achieve convergence-accelerated property prediction in MLPs. However, suppose % conservative + % exact greatly exceeds % mismatch. In that case, the SEM is likely to predict $\Delta k$ for training data that is small enough to accelerate the convergence of properties, yet large enough to limit the expenditure of computational resources on DFT calculations with unnecessarily small $\Delta k$.

Fig. 5 summarizes the analysis of the 4 SEMs applied to Al, Cu, and Mg. As shown in the figure, F-RMSE or V-RMSE has overall the highest likelihood across all three metals of predicting $\Delta k$ within the MLP convergence-accelerated regime. This observation is most evident in Al, while V-RMSE has the lowest % mismatch in Cu. AEAD, F-RMSE, and V-RMSE have equivalently low % mismatch in Mg. However, the relatively low % mismatch of F-RMSE and V-RMSE across all metals suggests that conservatively predicting $\Delta k_T^{MLP}$ with these SEMs is robust and will extend to other material systems. We note that utilizing SEMs to estimate $\Delta k$ for MLP training data within the convergence-accelerated regime in a novel material system requires making a selection which is based solely on the set of $\{\Delta k_T^{SEM}\}$ (highlighted in Fig. 4 and S3†) since $\{\Delta k_T^{MLP}\}$ (Fig. 3 and S2†) are unknown *a priori*. The most conservative approach is to select the minimum value of $\Delta k_T^{SEM}$ from the test dataset ($\{\Delta k_T^{SEM}\}_{min}$). For example, the V-RMSE has $\{\Delta k_T^{SEM}\}_{min} = 0.24$ Å$^{-1}$ for all three metals. Therefore, selecting the next smaller value of $\Delta k = 0.18$ Å$^{-1}$ for training data is likely to yield MLP models within the convergence-accelerated regime that predict material properties with nearly identical accuracy as models trained on $\Delta k = 0.12$ Å$^{-1}$ data. In fact, $\Delta k = 0.18$ Å$^{-1} < \Delta k_T^{MLP}$ for 100% of the MLP-predicted properties calculated by our models, showing that this conservative approach readily predicts $\Delta k$ within the convergence-accelerated regime. On the other hand, selecting $\Delta k = 0.24$ Å$^{-1}$ based on the F-RMSE $\{\Delta k_T^{SEM}\}_{min} = 0.30$ Å$^{-1}$ for Al and Cu results in $\Delta k < \Delta k_T^{MLP}$ for just 97.4% of properties for Al and Mg, but 100% for Cu, suggesting that the most conservative choice of $\Delta k$ (*i.e.*, the smallest $\{\Delta k_T^{SEM}\}_{min}$ among F-RMSE and V-RMSE) should be used without independently verifying $\{\Delta k_T^{MLP}\}$ for the material properties of interest. Regardless, this method to predict $\Delta k$ within the convergence-accelerated regime yields MLPs with nearly identical property prediction accuracy as models trained on $\Delta k = 0.12$ Å$^{-1}$, but with up to 7× reduction of the training dataset production cost due to using fewer $k$-points in the DFT calculations (see Table S1†).

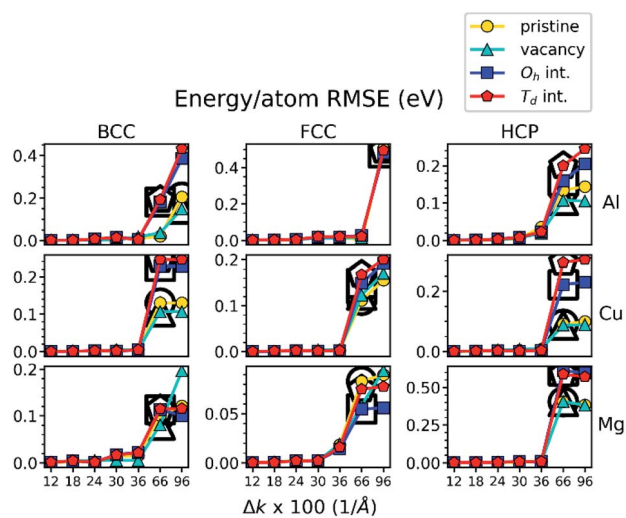

**Fig. 4** The E-RMSE of single-point DFT calculations as a function of $\Delta k$ for the 3600 configurations of the test dataset. Threshold $\Delta k$ ($\Delta k_T$) determined by the $\frac{1}{2}$MAD criterion is highlighted by prominent black-outlined symbols.

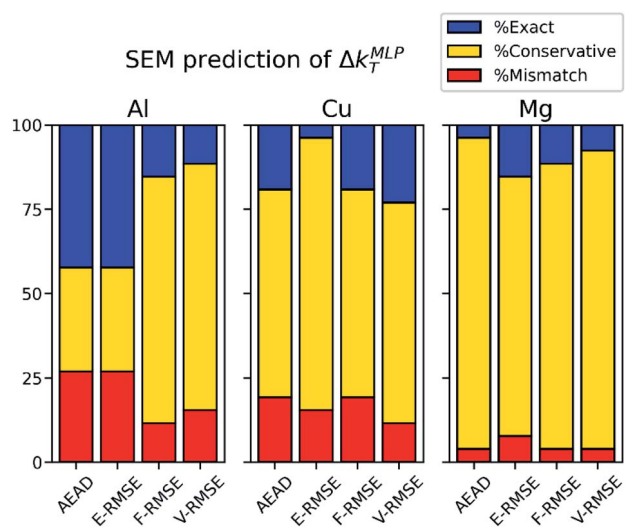

**Fig. 5** Analysis of SEM predictions of $\Delta k_T$ of MLP material properties ($\Delta k_T^{MLP}$). Proportions of % exact ($\Delta k_T^{MLP} = \Delta k_T^{SEM}$), % conservative ($\Delta k_T^{MLP} > \Delta k_T^{SEM}$), and % mismatch ($\Delta k_T^{MLP} < \Delta k_T^{SEM}$) over all properties, lattices, and structures is indicated for each SEM by blue, yellow, and red shading, respectively.

We suggest that SEMs can be used to predict the convergence acceleration regime of MLP material properties (Fig. 1) to reduce the computational cost of MLP production. The procedure is as follows: construct a reasonable set of structure samples from the anticipated material configuration space (analogous to the 3600 configurations in our DFT test dataset) by some inexpensive method. Then perform single-point DFT calculations on the set of structures over a range of $\Delta k$ from a small ultimate value up to a maximum value, such as that corresponding to gamma-point-only sampling in 2 × 2 × 2 supercells, as done in this work. Last, calculate SEMs of the DFT results, such as F-RMSE and/or V-RMSE, and their $\Delta k_T^{SEM}$ values based on a threshold criterion such as the $\frac{1}{2}$ MAD criterion (eqn (1)) utilized in this work. The $\Delta k$ to be used for training data is then selected based on the set $\{\Delta k_T^{SEM}\}$, for example, to be below the minimum value (the most conservative choice) or by some other method such as being below the mode (most frequent) value. In this work, we found that for Al, Cu, and Mg bulk metals in BCC, FCC, and HCP lattices, training with $\Delta k = 0.18$ Å$^{-1}$ guarantees nearly identical MLP property predictions as training with $\Delta k = 0.12$ Å$^{-1}$ due to the convergence acceleration effect, and $\Delta k$ can even be increased to 0.24 Å$^{-1}$ with no discernible impact on property prediction accuracy in Cu.

Our careful numerical investigations strongly indicate that the observed convergence acceleration of property predictions is due to the insensitivity of the MLP training process to random numerical noise introduced by reduced precision in the DFT training data if the noise remains below a system-and-property-dependent threshold. We hypothesize that convergence acceleration below threshold $\Delta k$ is a general effect in neural network-based MLPs trained on energy, force, and virial information. We further hypothesize that convergence acceleration is related to the empirical phenomenon well-known in the neural network research community of noisy training data improving neural network generalizability without harming accuracy.[51] We note that this connection implies that MLPs trained on larger $\Delta k$ within the convergence-accelerated regime are perhaps more generalizable than those trained on smaller $\Delta k$, which is a potential subject of future investigation. Furthermore, we suspect that similar findings can be expected with other convergence parameters typically employed in DFT calculations such as the size of the basis set used to expand the Kohn–Sham orbitals. In any case, this work demonstrates a method for utilizing the convergence acceleration effect to expedite MLP production by reducing computational resource consumption without sacrificing model accuracy.

## Conclusions

In this work, we conducted a quantitative study of the impact of the DFT calculation precision on material property prediction with MLPs trained on DFT datasets. We controlled precision with the $k$-space sampling density ($\Delta k$) and constructed parametrically identical training datasets of equal size and statistically equivalent sampling of the material configuration space to isolate the effect of $\Delta k$ on our MLPs. Under these controlled conditions, we identified a surprisingly robust stability in MLP property prediction as the DFT precision is reduced ($\Delta k$ increased) in the training dataset. We applied a scale-invariant criterion to define threshold values of $\Delta k$ below which MLP-predicted properties vary negligibly, and convergence is effectively accelerated with respect to $\Delta k$. On the other hand, DFT property predictions vary considerably with $\Delta k$, offering little guidance towards the selection of $\Delta k$ at which material property predictions are uniformly converged. We then showed that statistical properties of DFT data derived from energies, forces, and virials can in principle be utilized to predict the convergence-acceleration threshold $\Delta k$ without advance knowledge of the MLP convergence behavior of a material system. Finally, we demonstrated a method for determining $\Delta k$ for MLP training that leverages property convergence acceleration to reduce the expenditure of computing resources on training data production without sacrificing property prediction accuracy.

## Data availability

The code for MLP training can be found at https://github.com/deepmodeling/deepmd-kit and the code for property calculations can be found at https://github.com/deepmodeling/dpgen.

## Author contributions

Dylan Bayerl: conceptualization, data curation, formal analysis, investigation, methodology, resources, software, validation, visualization, writing – original draft. Christopher M. Andolina: formal analysis, investigation, visualization. Shyam Dwaraknath: discussed results and reviewed manuscript. Wissam A. Saidi: conceptualization, formal analysis, funding acquisition, investigation, methodology, project administration, resources, software, supervision. All authors discussed the results and reviewed the manuscript.

## Conflicts of interest

There are no conflicts to declare.

## Acknowledgements

We are grateful to the U.S. National Science Foundation (Award No. CSSI-2003808). This work was intellectually supported by the U.S. Department of Energy, Office of Science, Office of Basic Energy Sciences, Materials Sciences and Engineering Division under contract no. DE-SC0012704 and DE-AC02-05CH11231 (Materials Project program KC23MP). Also, we are grateful for computing time provided in part by the Pittsburgh Center for Research Computing (CRC) resources at the University of Pittsburgh, Extreme Science and Engineering Discovery Environment (XSEDE), which is supported by the National Science Foundation (NSF OCI-1053575), and Argonne Leadership Computing Facility, which is a DOE Office of Science User Facility supported under Contract DE-AC02-06CH11357.

## Notes and references